\documentclass[twocolumn,aps,pre,amsmath,showpacs]{revtex4-1}
\usepackage{graphicx,dcolumn,bm,amssymb,amsmath,ulem,indentfirst,amsthm}
\usepackage[toc,page]{appendix}
\usepackage{color}

\theoremstyle{plain}
\def\be{\begin{equation}}
\def\ee{\end{equation}}

\newtheorem*{theorem*}{Theorem}

\begin{document}

\author{Rico Milkus$^{1}$ and Alessio Zaccone$^{1,2}$}
\affiliation{${}^1$Statistical Physics Group, Department of Chemical
Engineering and Biotechnology, University of Cambridge, New Museums Site, CB2
3RA Cambridge, U.K.}
\affiliation{${}^2$Cavendish Laboratory, University of Cambridge, JJ Thomson
Avenue, CB30HE Cambridge,
U.K.}
\begin{abstract}
Viscoelasticity has been described since the time of Maxwell as an
interpolation of purely viscous and purely elastic response,
but its microscopic atomic-level mechanism in solids has remained elusive. We
studied three model disordered solids: a random lattice, the bond-depleted fcc
lattice, and the fcc lattice with vacancies. Within the harmonic approximation
for central-force lattices, we applied sum-rules for viscoelastic response
derived on the basis of non-affine atomic motions. The latter motions are a
direct result of local structural disorder, and in particular, of the lack of
inversion-symmetry in disordered lattices. By defining a suitable quantitative
and general atomic-level measure of nonaffinity and inversion-symmetry, we show
that the viscoelastic responses of all three systems  collapse onto a master
curve upon normalizing by the overall strength of inversion-symmetry breaking
in each system.  Close to the isostatic point for central-force lattices,
power-law creep $G(t)\sim t^{-1/2}$ emerges as a consequence of the interplay
between soft vibrational modes and non-affine dynamics, and various analytical
scalings, supported by numerical calculations, are predicted by the theory.

\end{abstract}

\pacs{}
\title{Atomic-scale origin of dynamic viscoelastic response and creep in disordered solids}
\maketitle
\section{Introduction}
The viscoelasticity of solids has been the object of intense debate at least
since the time of Maxwell. Continuum mechanics and relaxation models have
flourished through all the last century, with many extensions proposed to
capture different behaviours observed in metallurgy~\cite{Zener,Batist}.
For crystals with line defects, Andrade creep (whereby the relaxation shear
modulus presents the power-law scaling $G(t)\sim t^{-1/3}$) has been
convincingly explained
by Nabarro, Mott and others in terms of dislocation
dynamics~\cite{Mott,Nabarro}. Internal friction, which represents the imaginary
part of the viscoelastic response also known as the loss modulus $G''$, has
been interpreted in earlier models, in terms of diffusive motion of atoms
associated with defect mobility.

In glasses the situation is more complicated, because dislocations are
difficult to identify, and the origin of internal friction and complex
relaxation behaviour observed typically (power-law or stretched-exponential)
has remained unexplained. A recent work~\cite{Schirmacher} has applied elegant
field-theoretic methods within coherent-potential approximation, starting from
the assumption of spatially heterogeneous static shear modulus, to successfully
recover the $\alpha$-wing asymmetry in the resonance peak of $G''$ in
oscillatory rheology observed in experiments. However, the theory is on the
continuum level, and does not clarify which microscopic (atomic-level) features
ultimately control the viscoelastic response.

Recent simulation work~\cite{Samwer2014} motivated by this problem in the
context of metallic glasses, has shown that internal friction in glasses may
have its origin in quasi-localized correlated motions that have an
avalanche-like character. Furthermore, these excitations were found to be
suppressed in regions of high icosahedral symmetry.
Power-law creep $G\sim t^{-1/2}$ was recovered in previous work using mean
field theory~\cite{Yucht2013} and average stress fluctuations~\cite{Liu1996}.
The same result was found in a related field of athermal jammed solids, where
simulations and scaling arguments~\cite{Tighe1,Tighe2} based on Kelvin-Voigt
viscoelasticity have been combined with asymptotics of the vibrational density
of states (DOS) near the jamming transition (at which a jammed solid loses
rigidity) with average contact number $Z=6$, although the strongly non-affine
motion of the particles, which is crucial for disordered and jammed
solids~\cite{Barrat,Maloney}, was not explicitly taken into account in the
scaling analysis~\cite{Tighe1}. We improve on these methods by taking into
account the exact microstructure of the system as well as the non-affine
motions of all particles, providing a direct link between the microscopic
landscape and the frequency and time dependent shear modulus.

\section{Model Systems}
Here we re-examine this problem by considering three very different model
systems of amorphous solids in 3d, of which 2d slices are given in Fig.1.
We will work with a specific model of disordered harmonic spring networks
formed from the low-$T$ equilibration of dense Lennard-Jones fluids.
This is a good model for atomic disordered solids (defective crystals, metallic
glasses) but different from other types of disordered networks where
the preparation protocol may change the critical exponents and the critical
coordination numbers~\cite{Ellenbroek,Lubensky,Thorpe,Heussinger}.\\
The first lattice is a random network of harmonic springs generated according
to the protocol in Ref.~\cite{Milkus}: a Lennard-Jones glass is formed and
equilibrated in a metastable minimum, after which all nearest-neighbour
interactions are replaced by harmonic springs, all with the same spring
constant $\kappa$ and with a relatively narrow distribution of spring length
$R_{0}$. Upon randomly cutting the harmonic bonds in the sample, lattices with
variable coordination number $Z$ can be formed. In the present work this
depletion process is performed in such a way that we get a very narrow
distribution of coordination numbers to avoid effects stemming from fluctuating
connectivity in the system.\\
The two fcc lattices (the bond-depleted, Fig.1b, and with vacancies, Fig.1c)
are instead generated starting from a perfect fcc lattice with $Z=12$ and same
spring constant $\kappa$ and lattice constant $R_{0}$ as the random lattice.
The microstructure, and in particular the local symmetry, of the three lattices
is, however, very different. For example, in Ref.~\cite{Milkus} it was shown
that the standard bond-orientational order parameter $F_{6}$, which measures
the spread in the orientations of bonds on the lattice~\cite{Steinhardt}, is
practically equal to $1$ for the bond-depleted fcc (for any $Z$ value), whereas
it is much lower ($\simeq 0.3$) for the random lattice.\\
For these models we develop an analytical theory of viscoelastic response based
on the non-affine deformation formalism, which is a fully microscopic approach.
Our analysis shows that, surprisingly, the oscillatory moduli of these systems
fall onto a master curve after normalizing by an order parameter which
describes the average degree of local inversion-symmetry on any atom. The same
order parameter controls the non-affine particle rearrangements that have a
cooperative quasi-localized character, which explains the findings of
simulations~\cite{Samwer2014}. Further, the power-law creep $G\sim t^{-1/2}$
found near the isostatic transition of all the three lattices is shown to be
the consequence of both the excess of soft modes in the DOS, and crucially,
also of the underlying non-affine dynamics.
\begin{figure}
\begin{center}
\includegraphics[height=3.5cm,width=8.7cm]{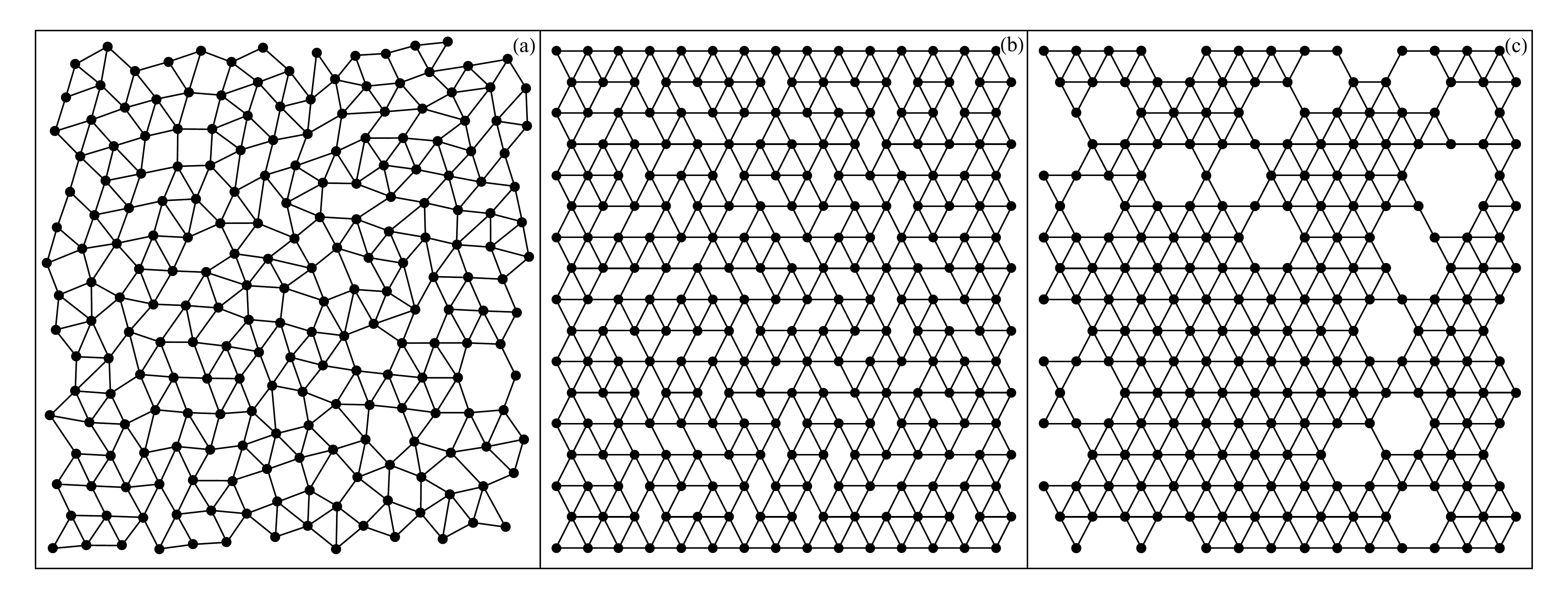}
\caption{Two dimensional schematic of our 3D model systems. (a) the random network,
(b) the fcc lattice with randomly cut bonds and (c) the fcc lattice with
randomly removed atoms.}
\end{center}
\end{figure}
\section{Formalism}
The starting point of our analysis is the microscopic equation of motion for a
particle in a disordered lattice, which was derived for the case of a
phenomenological damping
motion with constant damping coefficient $\nu$, in Ref.~\cite{Lemaitre} and was
shown, also in~\cite{Lemaitre}, to reduce to a simple harmonic-oscillator type
equation for the deviation variable $\underline{x}_{i}$ which measures the
particle displacement from the original position:
\begin{equation}
m \ddot{\underline{x}}_{i}+ \nu \dot{\underline{x}}_i  +
\underline{\underline{H}}_{i j}  \underline{x}_j = \underline{\Xi}_{i, \kappa
\chi} \eta_{\kappa \chi} .
\end{equation}
We used the Hessian of the system $\underline{\underline{H}}_{i j} = -
\partial^2 \mathcal{U}/\partial \underline{r}_i\partial \underline{r}_j = -
{\partial \underline{f}_i}/{\partial \underline{r}_j}$ and the non-affine force
$\underline{\Xi}_{i, \kappa \chi} = {\partial \uline f_i}/{\partial
\eta_{\kappa \chi}}$. Here, $\eta_{\kappa \chi}$ denotes the Cauchy strain
tensor for a generic deformation field. For a shear deformation, ${\kappa
\chi}\equiv {xy}$. The non-affine force $\uline \Xi_{i, \kappa \chi}$
represents the net force that acts on a particle that is en route towards its
affine position. If the particle's original position in the undeformed lattice
is
$\underline{R}_{0}$, the affine position is defined as
$\underline{r}_{i,A}=\underline{\underline{\eta}}\underline{R}_{i,0}$. In a
perfectly centrosymmetric lattice, the particle en route towards this affine
position receives forces from its nearest-neighbours which cancel each other
out by symmetry, leaving the particle at equilibrium in the affine position. In
a disordered lattice, due to local breaking of inversion-symmetry on the given
particle, these forces do not cancel, and their vector sum is a net force that
brings the particle to a final (non-affine) position which differs from
$\underline{r}_{i,A}$.
For a generic harmonic lattice with no pre-stress, the non-affine force vector
is defined as $\underline{\Xi}_{i, \alpha \beta} \,=\, - R_0 \kappa \sum_j
\hat{n}_{i j}^\alpha \hat{n}_{i j}^\beta \underline{\hat{n}}_{i j}$, with $R_0$
and $\kappa$ being the rest distance and force constant between the particles.
The sum is performed over the nearest neighbours and includes the unit bond
vector $\underline{\hat{n}}_{i j} $ pointing from atom $i$ to $j$.

Normal-mode decomposition of the terms in Eq.(1) onto the eigenvectors
$\underline{v}_p$ (where $p=1...N$) of the Hessian, and taking the Fourier
transform of the equation of motion as in Ref.\cite{Lemaitre}, lead to the
complex viscoelastic shear modulus for oscillatory shear deformation (with
imposed frequency $\Omega$):
\begin{equation}
        G^{*}(\Omega) \,=\,  G^{A} - 3 \,\rho \int_{0}^{\omega_{D}}
\frac{D(\omega) \Gamma(\omega)}{m \omega^2 - m \Omega^2 + \mathrm{i} \nu
\Omega} d \omega .
        \end{equation}
Here we introduced the frequency correlator of the non-affine forces,
$\Gamma_{xyxy}(\omega) = \langle \hat \Xi_{p, xy}\, \hat \Xi_{p, xy} \rangle_{p
\in \{ \omega, \omega + \delta \omega \}}$, where $\hat \Xi_{p, xy} =
\underline{\Xi}_{xy} \cdot \underline{v}_p$. Also, $\rho=N/V$ is the atomic
density, or number of atoms (or nodes) on the lattice per unit volume. $G^{A}$
is the affine shear modulus (also known as the Born-Huang modulus), which is
independent of the applied frequency $\Omega$, and coincides with the elastic
response in the limit $\Omega \rightarrow \infty$.
Here, $\omega$ denotes the eigenmode frequency of internal vibrations of the
lattice, and $\omega_{D}$ denotes the Debye frequency, i.e. the highest
frequency of the vibrational spectrum. The latter spectrum, i.e. the normalized
distribution of vibrational eigenmodes is represented by the DOS, denoted here
as $D(\omega)$. The mass of the particles $m$ is set to $1$ for the reminder of
the paper, since it's of no concern in the present work.\\
The above sum-rule allows the calculation of the complex shear modulus for any
harmonic lattice for which both the DOS and the correlator function
$\Gamma(\omega)$ can be easily evaluated numerically. For the DOS we follow the
same procedure as in Ref.~\cite{Milkus}, whereas for $\Gamma(\omega)$ we follow
the procedure of Ref.~\cite{Lemaitre}. This is a straightforward exercise for
the three model lattices shown in Fig.1. \\

\section{Results}
\begin{figure}[h]
                \includegraphics[height=7.9cm,width=8.5cm]{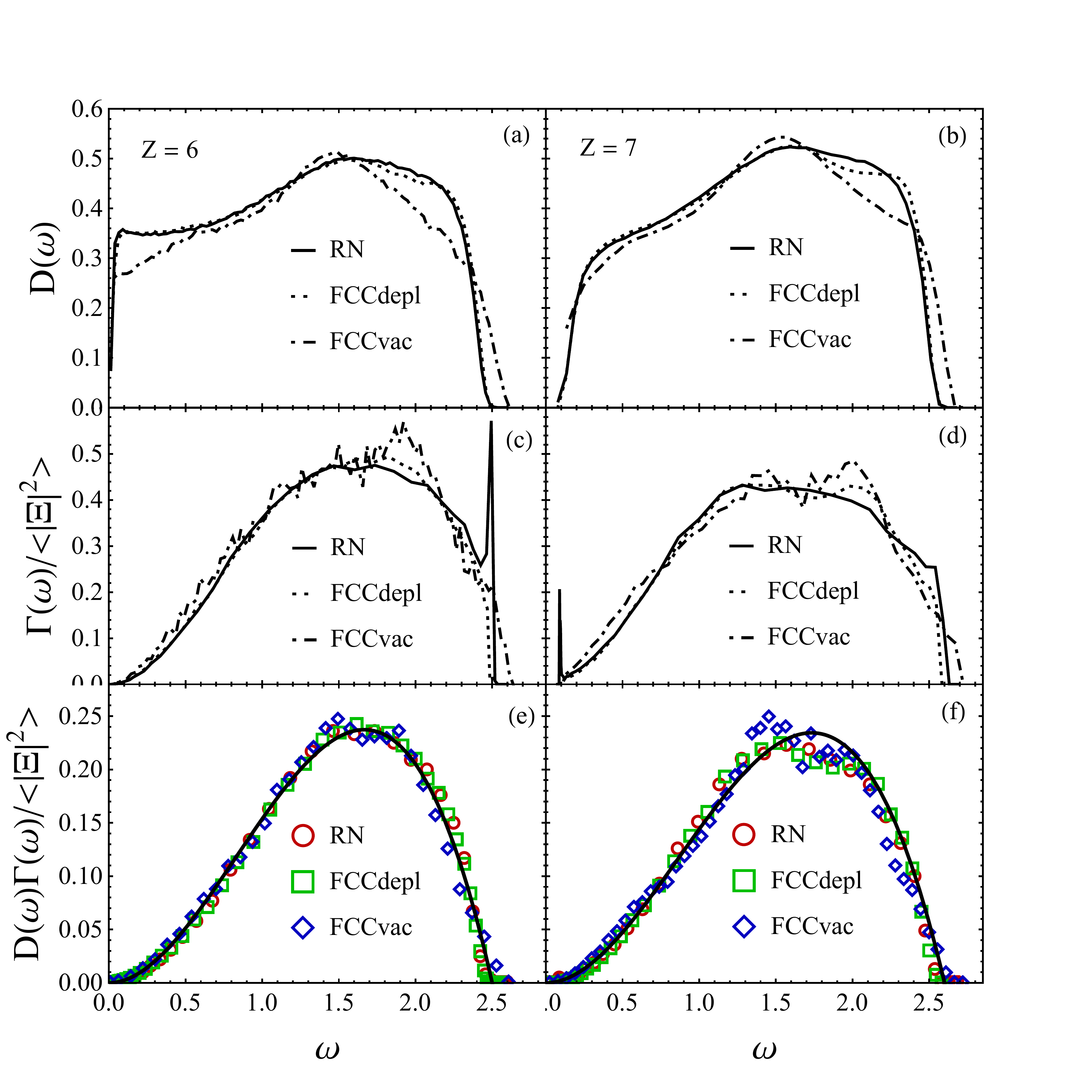}
                \caption{Density of states $D(\omega)$ (a,b) and correlator
function $\Gamma(\omega)$ (c,d) for the three different model systems and two
different coordination numbers ($Z=6.1$ and $Z=7$, respectively). They are
normalized by $\langle \vert \underline{\Xi} \vert^{2}\rangle / \rho$, proportional to
the average absolute square of the non-affine force field to obtain a master
curve. (e,f) show the product $D(\omega)\cdot\Gamma(\omega)$ which appears in
the formula for $G^{*}(\Omega)$. Remarkably, the product of the two functions
can be conveniently fitted by a simple cubic function $\omega^2 (\omega_D -
\omega)$, represented as a solid line.}\label{fig:DOSGamma}
        \end{figure}

We have calculated $G^{*}(\Omega)$ for the three lattices with two
different average
coordination numbers $\left\langle Z \right\rangle$: $Z =7.0$, where
all lattices are mechanically well stable; $Z = 6$ (for the fcc with
vacancies) and $Z = 6.1$ (for the two bond-depleted systems), i.e.
very close to the point of marginal stability. First we calculated the
vibrational
density of states $D(\omega)$ and the correlator function $\Gamma(\omega)$,
which are shown in Fig. \ref{fig:DOSGamma}. Since these quantities appear as
the $D(\omega)\Gamma(\omega)$ product in Eq.(2), it is convenient to study this
product as a single function of $\omega$.

Remarkably, we notice from Fig. \ref{fig:DOSGamma} that, although $D(\omega)$
and $\Gamma(\omega)$ behave differently for each of the three systems and have
a rather complicated form, their product, when normalized by the quantity
$\langle \vert \underline{\Xi} \vert^{2}\rangle / \rho$, shows a strikingly
\textit{universal} behaviour over the full frequency range, and can be fitted
by a simple cubic function of $\omega$, of the form
        \begin{equation}\label{eq:Fitting}
        \frac{D(\omega)\Gamma(\omega)}{\langle \vert \underline{\Xi}
\vert^{2}\rangle / \rho} \,\sim\, \omega^2 (\omega_D - \omega).
        \end{equation}
Here, the quantity $\langle \vert \underline{\Xi} \vert^{2}\rangle $ is
evaluated by taking the square of the absolute value of each vector $\vert
\underline{\Xi}_{i} \vert$, constructed for each atom $i$, and averaging over
all atoms in the system. This same quantity has been used to form a suitably
normalized order parameter in~\cite{Milkus}.

Since $D(\omega)$ approaches a low-$\omega$ plateau in the limit of marginal
stability ($Z\rightarrow 6$), as is known from many studies in the
past~\cite{OHern, Silbert}), the low frequency behaviour of
$D(\omega)\cdot\Gamma(\omega)$ is dominated by the correlator function
$\Gamma(\omega) \sim \omega^2$, a result that was derived in
Ref.~\cite{Zaccone2011}. It is interesting to note Dirac-delta spikes in
$\Gamma(\omega)$, which happen at frequencies that correspond to
strongly-localized modes: at $Z=6$ a spike is visible near the top of the
spectrum, where modes tend to be Anderson-localized. At $Z=7$, instead, a spike
is visible at a frequency close to the Ioffe-Regel crossover~\cite{Tanaka} (and
to the boson peak frequency) where modes are also strongly
localized~\cite{Milkus}.

\begin{figure}[h]
                \includegraphics[trim=0cm 1cm 3cm 3cm,clip=true,
height=7.1cm,width=8.5cm]{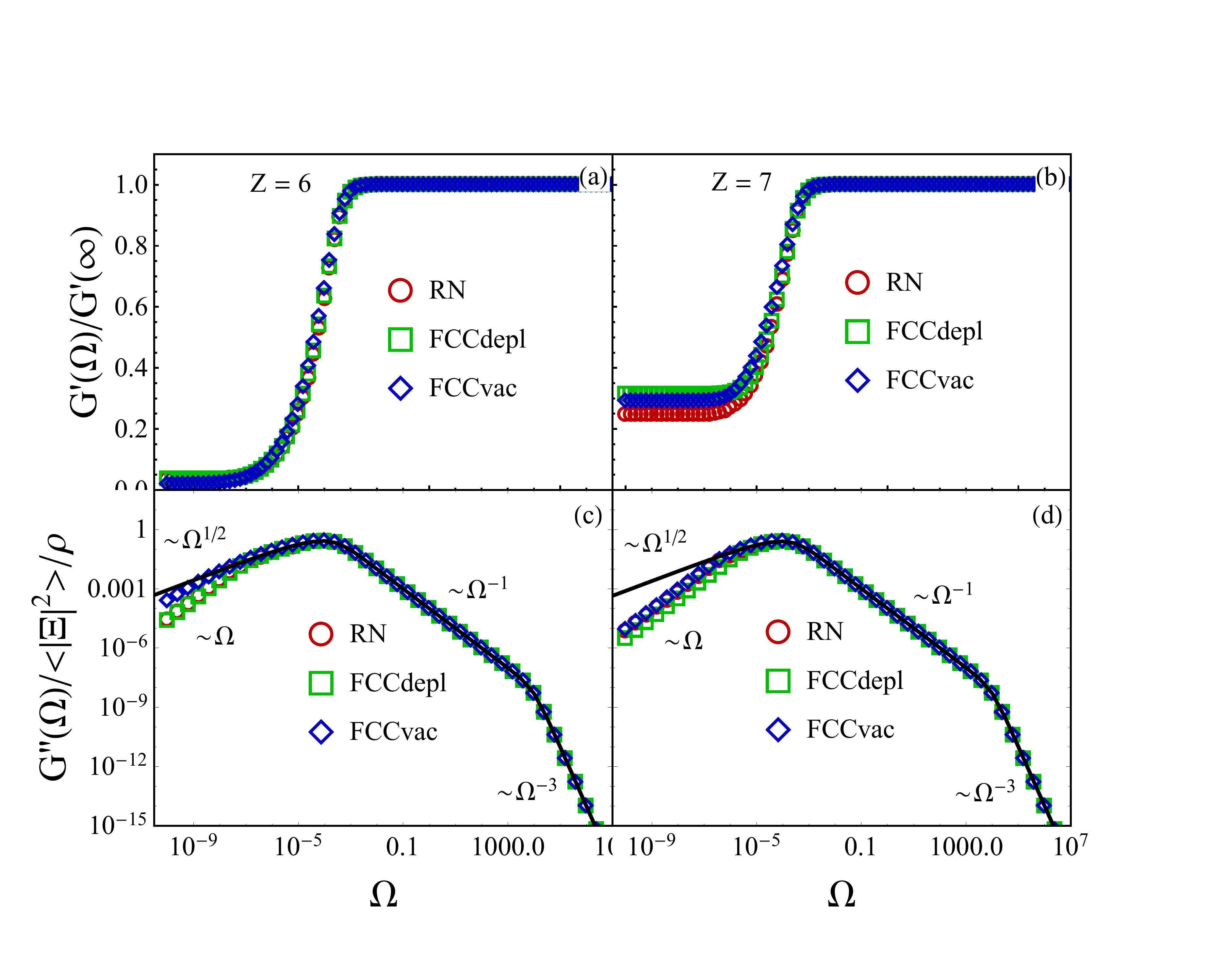}
                \caption{$G'$ (a,b) and $G''$ (c,d) of our three model systems
for $Z=6$ (left) and $Z=7$ (right), respectively. In (c,d) in order to collapse
the loss modulus of the three systems onto a single master curve, we have
normalized by the factor $\langle \vert \underline{\Xi} \vert^{2}\rangle /
\rho$.}\label{fig:GW}
        \end{figure}

Let us now consider the viscoelastic response of the three models systems. We
use the convention of splitting the complex shear modulus into its real and
imaginary part $G(\Omega) = G'(\Omega) + \mathrm{i} G''(\Omega)$. Both moduli
can be calculated according to
        \begin{eqnarray}
        G'(\Omega) &=& G^{A} - 3  \rho \int_{0}^{\omega_D}
\frac{D(\omega)\Gamma(\omega) (\omega^2 - \Omega^2)}{(\omega^2 - \Omega^2)^2 +
\nu^2 \Omega^2} \\
        G''(\Omega) &=& 3  \rho \int_{0}^{\omega_D}
\frac{D(\omega)\Gamma(\omega)\nu \Omega}{(\omega^2 - \Omega^2)^2 + \nu^2
\Omega^2},
        \end{eqnarray}
and are plotted in Fig.3. In the numerical calculation we implemented the
convenient cubic form of the product $D(\omega)\Gamma(\omega)$ that was shown
above to be an excellent fitting to the numerical evaluation of these
functions. Various scalings have been reported in the plots, which can be
extracted from the asymptotic analysis of Eq.(4)-(5). Most notable of
which are the low frequency scalings scalings of $G''(\Omega)$, which agree
very well with EMT results from Ref.~\cite{Yucht2013} ($G''\sim \Omega^{1/2} $
for $ Z\approx 6$ and $G''\sim \Omega$ for $ Z = 7$). Deviations from their
numerical results ($G''\sim \Omega^{0.41}$) might be caused by finite
temperature effects in their simulations (whereas our calculation is carried
out at $T=0$). In our work we have used
different values of the damping coefficient $\nu$ to study its influence on the
results. We found that it has no influence on the qualitative behaviour of $G'$
and $G''$, besides when it approaches very small values, where we get divergent
results. Aside from that $\nu$ only shifts the values to smaller $\Omega$ and
expands the range of the $\sim \Omega^{-1}$ scaling in $G''$. We therefore
chose a quite large value $\nu = 10'000$ to demonstrate this behaviour clearly
and to focus on the physically important case of overdamped dynamics typical of
amorphous solids (metallic glasses, organic glasses, foams etc).

\begin{figure}[h]
        \centering
        \includegraphics[trim=0cm 3cm 0cm 2cm,
clip=true,height=4.1cm,width=9.1cm]{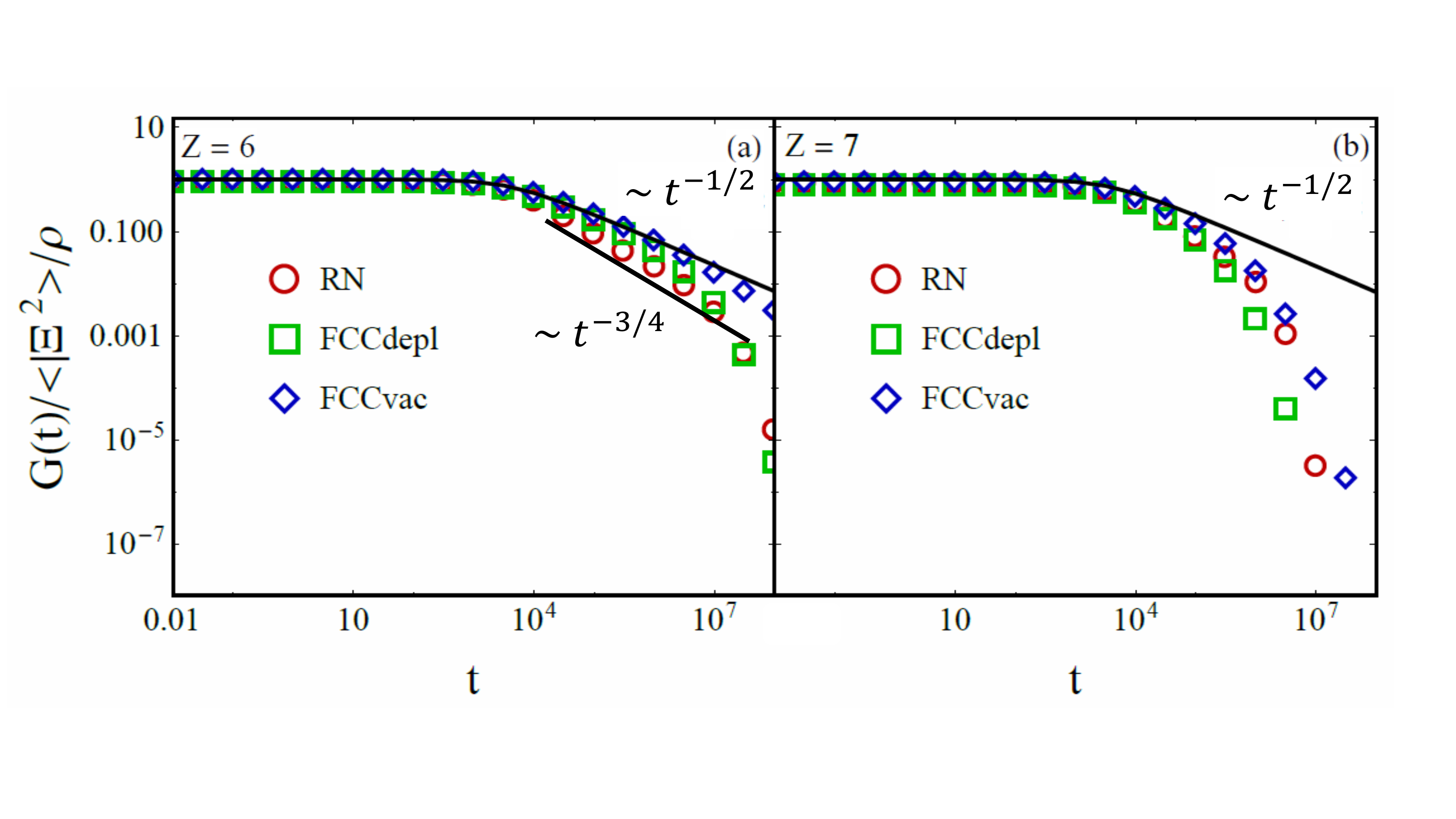}
        \caption{$G(t)$ of our three model systems for $Z=6$ (left) and $Z=7$
(right). We have indicated the different scaling ranges. One can see that the
power-law scaling, which is present on $3-4$ orders of magnitude in time for
$Z=6$, breaks down for $Z=7$. This is due to due to the expanding Debye $\sim
\omega^{2}$ regime in the DOS, see text.}\label{fig:Gt}
    \end{figure}

Next we consider the time-dependent shear modulus $G(t)$, which can be
calculated by taking the inverse Fourier transform of $G^{*}(\Omega)$,
\begin{equation}
\begin{split}
G(t)&=G- \frac{3}{2\pi} \rho \int_{-\infty}^{\infty} \int_{0}^{\infty} \frac{D(\omega)
\Gamma(\omega) \exp(\mathrm{i} \Omega t)}{\omega^2 - \Omega^2  + \mathrm{i} \nu
\Omega} d\omega d\Omega\\
&= G - 3 \rho\,t e^{- \frac{\nu}{2}t} \int_{0}^{\infty} D(\omega)
\Gamma(\omega) \,\mathrm{sinc}(\frac{1}{2} \sqrt{4 \omega^2 - \nu^2} t)
d\omega.
\end{split}
\end{equation}
Here $G$ is the quasistatic (infinite-time or zero-frequency) shear modulus (which has a strong nonaffine character), $\mathrm{sinc}(x) = \sin(x)/x$ denotes the cardinal sine function.
Numerical evaluation of Eq.(6) for the three lattices at the two representative
values of $Z$ are reported in Fig.4. Again we took advantage of the simple
cubic fitting Eq.(3) for the product $D(\omega)\Gamma(\omega)$, which allows
one to avoid the problem of a numerical gap between zero frequency and the
first eigenfrequency (this gap is not negligible for systems with $N<10^{5}$
and our simulated lattices have $N=5\times 10^{4}$). For small times we observe
a plateau that corresponds to the high-frequency affine response, after which a
power-law decay is observed with an exponent comprised in the range between
$-1/2$ and $-3/4$. This power-law can be understood mechanistically as follows.

We focus on the limit overdamped systems, which is both important and turns to
be amenable to analytic simplifications.  For large $\nu$ and large times we
can simplify the expression in Eq.(6). First we take $\sqrt{\nu^2 - 4 \omega^2}
\approx \nu - 2\,\frac{\omega^2}{\nu}$, where we use $\omega \ll \nu$.
 We insert this into Eq.(6) and use the definition of $\sinh(x)$ to get
    \begin{equation}\label{eq:Gtapprox}
    \begin{gathered}
    G(t)\,\approx\,6 \rho\, e^{- \frac{\nu}{2}t} \int_{0}^{\infty}
\frac{D(\omega) \Gamma(\omega) \sinh(\frac{\nu}{2} t - \frac{\omega^2}{\nu}
t)}{\omega^{2}(\nu - 2\frac{\omega^2}{\nu})} d\omega\\
    =\, 3 \rho\, \int_{0}^{\infty} \frac{D(\omega) \Gamma(\omega)(e^{-
\frac{\omega^2}{\nu} t} - e^{-\nu t + \frac{\omega^2}{\nu} t})}{\omega^{2}(\nu
- 2\frac{\omega^2}{\nu})} d\omega\\
    \approx\, 3 \rho\,\frac{1}{\nu} \int_{0}^{\infty} \frac{D(\omega)
\Gamma(\omega)}{\omega^{2} } e^{- \frac{\omega^2}{\nu} t} d\omega.
    \end{gathered}
    \end{equation}
    In the last step we have used $\nu \gg 2 \omega^2/\nu$ and $\nu t -
\omega^2 t/\nu \gg 1$. This corresponds to a system of Maxwell elements with
relaxation times $\tau = \nu/\omega^2$. We now recall the standard relationship
between the DOS and the eigenvalue spectrum $\rho(\lambda)$ of the Hessian
matrix, $D(\omega)d\omega = \rho(\lambda)d\lambda$, with $\omega^{2}=\lambda$.
At the isostatic point of disordered solids, $Z=6$ the DOS develops a plateau
of soft modes, which is visible in our Fig.2(a). This limit corresponds to the
scaling $\rho(\lambda) \sim \lambda^{-1/2}$ in the eigenvalue distribution,
which arises from the dominance of random-matrix behaviour in the spectrum, and
this scaling can be derived e.g. from the famous Marcenko-Pastur distribution
of random-matrix theory, as discussed recently in ~\cite{Parisi}. In our DOS, a
scaling $\rho(\lambda) \sim a + \lambda^{-1/2}$, where $a$ is a constant, is
more appropriate since we are in fact slightly above $Z=6$, and this will
explain the power-law exponents in $G(t)$ larger than $1/2$ in our
calculations. However, we will stick to the simple $\rho(\lambda) \sim
\lambda^{-1/2}$ for the asymptotic analysis. Recall now that
$\Gamma(\omega)\sim\omega^{2}$, from the analytical theory of non-affine
deformations~\cite{Zaccone2011}, which implies
$\tilde{\Gamma}(\lambda)\sim\lambda$. Inserting these results in the last line
of Eq.(7) we obtain the following Laplace transform which can be easily
evaluated to:
    \begin{equation}
    \begin{split}
    G(t) & \sim \int_{0}^{\infty}
\frac{\rho(\lambda)\tilde{\Gamma}(\lambda)}{\lambda} e^{- \lambda t} d\lambda
\sim \int_{0}^{\infty} \frac{\lambda^{-1/2}\lambda}{\lambda} e^{- \lambda t}
d\lambda\\
    & \sim t^{-1/2}.
    \end{split}
        \end{equation}

This scaling for the power-law creep modulus was shown in
simulations of creep in athermal jammed systems in Ref.~\cite{Tighe1}, using a
system of Kelvin-Voigt elements (whereas we use a standard-linear-solid or
Zener material). The theoretical argument that was proposed to
explain the scaling $t^{-1/2}$ was not fully microscopic, because the
correlator between eigenmodes and shear field 
was taken to be independent of the eigenfrequency, hence constant on average
for a given frequency interval. This is not a physically justified approximation, because
the correlator $\Gamma(\omega)$ in our data (and also in Ref.\cite{Lemaitre}) displays a strong (and non-random) dependence on the eigenfrequency as one can
see in Fig. 2 (c,d). Our model
improves substantially on this aspect, by including the eigenfrequency dependence of the
non-affine
correlator into the theoretical analysis of the scaling. In this way, our framework provides a direct
link between the
microscopic nonaffine dynamics and the viscoelastic moduli.

\section{Conclusion}
Using the non-affine response formalism, we studied three model harmonic
lattices with disorder, which have very different microstructure (as reflected
in e.g. different values of bond-orientational order parameter as shown in
previous work~\cite{Milkus}). Yet, the three different lattices have
qualitatively the same (universal) viscoelastic response, i.e. $G'$ and $G''$
collapse onto master curves as a function of frequency, once the moduli are
normalized by a factor $\langle \vert \underline{\Xi} \vert^{2}\rangle / \rho$,
where $\rho$ is the atomic density. Here $\langle \vert \underline{\Xi}
\vert^{2}\rangle$ is crucially related to the symmetry that controls this
universality: the local degree of inversion-symmetry. This is evident from the
definition of the non-affine force vector for harmonic lattices:
$\underline{\Xi}_{i, xy} = - R_0 \kappa \sum_j \hat{n}_{i j}^x \hat{n}_{i j}^y
\underline{\hat{n}}_{i j}$. The norm of this vector is clearly identically zero
for all atoms in a perfectly centrosymmetric lattice with no defects, whereas
its value is larger for lattices where the inversion-symmetry is lowered.
Hence, the parameter $\langle \vert \underline{\Xi} \vert^{2}\rangle$ crucially
is proportional to the overall (spatially-averaged) degree to which local
inversion symmetry is broken in a disordered lattice.

These results thus identify the atomic-scale origin of internal friction and
viscoelastic response in amorphous solids (e.g. glasses) with the local
inversion-symmetry breaking, which is the same effect that causes a softer
elastic response~\cite{Milkus} and is associated with quasi-localized
avalanche-like non-affine motions~\cite{Harrowell}. Our framework provides a
clear theoretical explanation to recent simulations results~\cite{Samwer2014}
where internal friction was shown to correlate with cooperative non-affine
motions and regions of lower local symmetry. This framework will play an
important role for the rational design of new materials with tailored
viscoelastic response and energy absorption properties in many materials
science and engineering applications.

\end{document}